# Distribution-Based Sub-Population Selection (DSPS): A Method for in-Silico Reproduction of Clinical Trials Outcomes

*Mohammadreza Ganji, Anas El Fathi Ph.D., Chiara Fabris Ph.D., Dayu Lv Ph.D., Boris Kovatchev Ph.D., Marc Breton Ph.D.* [1]

**Abstract**

*Background and Objective:* Diabetes presents a significant challenge to healthcare due to the negative impact of poor blood sugar control on health and associated complications. Computer simulation platforms, notably exemplified by the UVA/Padova Type 1 Diabetes simulator, has emerged as a promising tool for advancing diabetes treatments by simulating patient responses in a virtual environment. The UVA Virtual Lab (UVLab) is a new simulation platform to mimic the metabolic behavior of people with Type 2 diabetes (T2D) with a large population of 6062 virtual subjects. The objective of this work is to propose a statistical method for selecting a subset from this large pool of virtual subjects, such that the selected group possesses the desired features necessary to reproduce and predict the outcomes of a clinical trial.

*Methods:* The work introduces the Distribution-Based Population Selection (DSPS) method, a systematic approach to identifying virtual subsets that mimic the clinical behavior observed in real trials. The method transforms the sub-population selection task into a Linear Programing problem, enabling the identification of the largest representative virtual cohort. This selection process centers on key clinical outcomes in diabetes research, such as HbA1c and Fasting plasma Glucose (FPG), ensuring that the statistical properties (moments) of the selected virtual sub-population closely resemble those observed in real-word clinical trial. Notably, the method is adaptable to specific research goals, allowing the fine-tuning of sample sizes and acceptable margins of error.

*Results:* DSPS method was applied to the insulin degludec (IDeg) arm of a phase 3 clinical trial, which evaluated the efficacy and safety of the combination therapy of insulin degludec and the GLP-1 receptor agonist liraglutide, compared to each component administered alone. This method was used to select a sub-population of virtual subjects that closely mirrored the clinical trial data across multiple key metrics, including glycemic efficacy, insulin dosages, and cumulative hypoglycemia events over a 26-week period. The relative sum of square errors between the selected sub-population and the clinical trial data's mean and standard deviation was 0.33, and the percentage error was 1.07% (SD =1.2%).

*Conclusion:* The DSPS algorithm is able to select virtual sub-population within UVLab to reproduce and predict the outcomes of a clinical trial. This statistical method can bridge the gap between large population simulation platforms and previously conducted clinical trials, providing researchers virtual sub-populations with the specific properties required for targeted studies.

---

[1] Corresponding author, mb6nt@virginia.edu

1. Introduction

Diabetes is a chronic disorder affecting the body's ability to control blood sugar levels[1], [2]. Both Type 1 and Type 2 diabetes present unique challenges; Type 1 diabetes (T1D) is an autoimmune disorder that causes insufficient insulin production, while Type 2 diabetes (T2D) is frequently associated with insulin resistance [3], [4]. Careful management of diabetes is necessary to prevent complications, including kidney disease, heart disease, microvascular damage and neuropathy[5], [6], [7]. Given these challenges, innovative solutions, such as digital twin technology, are increasingly being explored to enhance diabetes management and treatment outcomes[8].

Digital twins, virtual representations of real individuals, have gained increasing attention in healthcare for their potential applications[9]. Within the field of diabetes, with a particular focus on T1D, digital twins have played a central role in the advancement of treatment approaches[10], [11], [12]. These platforms not only provide researchers with the means to evaluate their novel interventions in a virtual environment, thereby circumventing the need for preliminary costly clinical trials, but they have also been instrumental in designing model-based dosing strategy algorithms[13], [14], [15], [16], [17], [18], [19]. The UVA/Padova T1D simulator, a simulation platform accepted by the U.S. Food and Drug Administration (FDA), has been utilized for years as a substitute for animal pre-clinical trials, opening new possibilities for designing personalized treatments[20], [21], [22], [23], [24], [25]. Encouraged by these successes in T1D, similar efforts have been directed toward developing and validating simulation platforms for T2D, which presents its own distinct challenges and complexities.

More recently, efforts have been made to validate the simulation platforms that can mimic the metabolic behavior of patients with T2D. Visentin et al. have developed the Padova T2D with 100 virtual subjects representing individuals with different stages of T2D [26]. An expanded version of Hovorka's T1D simulator has been adapted to model multiscale glycemic dynamics specific to T2D [27], [28]. At the University of Virginia, a simulation platform called the UVa Virtual Lab (UVLab) has been developed, consisting of 6062 subjects to represent subtypes of people with T2D at different stages. This platform is capable of executing a clinical trial protocol and replaying a specific clinical trial in silico. While the UVLab, with its 6,062 virtual subjects, can execute large in-silico experiments, fully leveraging this comprehensive platform requires an understanding of how to select sub-populations that mirror the diverse metabolic behaviors observed in specific groups of real-world patients.

To effectively tailor UVLab's capacity to a specific scientific need, it is crucial to select a sub-population that provides the glycemic behavior required for the specific application under a specific clinical protocol. This process begins with executing the same clinical protocol within the simulation platform and defining the glycemic metrics of interest. These glycemic metrics can be represented by their measures of central tendency and variability (moments of distribution), typically expressed as the mean and variance in the context of diabetes-related clinical trials. Given the importance of accurately replicating clinical outcomes, there is a need for a systematic approach to select virtual sub-populations that closely mirror the clinical behavior observed in real-world trials. The goal of this paper is to propose a statistical method for identifying such virtual subsets, thereby enhancing the ability of simulation platforms to reproduce and predict clinical outcomes for patients with T2D. To demonstrate this method, a specific clinical trial was chosen; however, the scope of application of this method is not limited to any particular clinical trial [30].

The following sections outline the DSPS methodology and its application to reproduce and predict clinical trial outcomes within the UVLab. The results will demonstrate the method's accuracy in selecting virtual cohorts to reproduce and forecast outcomes for the insulin degludec (IDeg) arm of a phase 3 clinical trial.

## 2. Material and Methods

Let $\mathcal{P} = \{e_1, e_2, \ldots, e_{n_p}\}$ represent the parent in-silico population, where each $e_i$ is a simulated member of the population. And let $f = \{f_1, f_2, \ldots, f_{n_X}\}$ represent the features of interest, where $n_X$ denotes the number of features. For instance, this set could represent features such as Hemoglobin A1c (HbA1c) and FPG at baseline, or the change in CGM measured time in range between baseline and study week 26. Let $\mathbb{D} = \{x_i | i \leq n_D, x_i \in \mathbb{R}^{n_X}\}$ represent the dataset where each row $x_i$ corresponds to the features of a member $e_i$ from the parent population $\mathcal{P}$.

The goal is to identify $\mathcal{S} \subseteq \mathcal{P}$ such that the sample moments of the features in $\mathcal{S}$ match a set of target criteria $T = \{(f_i, k_i, t_i) | i \leq n_T, f_i \leq n_x, k \in \mathbb{N}\}$. Here, each element $(f_i, k_i, t_i)$ represents a feature $f_i$, the $k_i$-th sample moment, and its target value $t_i$. $M_{f_i}^{k_i}$ denotes the target for $k_i$-th moment of the $f_i$ feature.

### 2.1. Distribution-Based Sub-Population Selection (DSPS)

DSPS is a systematic sub-population selection method that involves selecting a sample $\mathcal{S}$ from a population $\mathcal{P}$ based on their distribution characteristics of specific features of interest $f$. The problem of sub-population selection can be formulated as a Bernoulli process, where each member of the parent population $e_i$ has a certain probability $p_{e_i} \in [0\ 1]$ of being selected in the sample $\mathcal{S}$.

In this Bernoulli process, $p_{e_i}$ represents the probability of selection for each element $e_i$. $b_{e_i}$ is a binary indicator where $b_i = 1$ if the member $e_i$ is included in the sub-population $\mathcal{S}$ (i.e., $e_i$ is selected), and $b_i = 0$ otherwise. Given the uniformly sampled $r_i \in [0\ 1]$ for each $e_i$, the selection rule is defined as follows:

$$b_{e_i} = \begin{cases} 1 & if\ p_{e_i} > r_i \\ 0 & if\ p_{e_i} \leq r_i \end{cases}$$

This rule indicates whether element $e_i$ is present in the sample $\mathcal{S}$. The process thus relies on assigning a selection probability $p_{e_i}$ to each element $e_i$, determining their eligibility for inclusion in the selected sub-population.

A critical component of this Bernoulli process is determining the sample size $n_t$. The sample size can be predefined based on the application's requirements or remain flexible. The objective is to select a sample $\mathcal{S}$ where the moments of the distribution for specific features closely match the target moments.

To achieve this, the system of linear equations (SLE) of the form $A_{SLE} \cdot P_{SLE} = C_{SLE}$ (1), derived in Appendix 1, can be solved to obtain the values of $p_i$ with $0 \leq p_{e_i} \leq 1$ for all $i$.

$$A_{SLE} = \begin{bmatrix} 1 & 1 & \cdots & 1 \\ x_1^{f_1} & x_2^{f_1} & \cdots & x_{n_p}^{f_1} \\ \vdots & \vdots & \ddots & \vdots \\ x_1^{f_{n_X}} & x_2^{f_{n_X}} & \cdots & x_{n_p}^{f_{n_X}} \\ \left(x_1^{f_1} - M_{f_1}^1\right)^2 & \left(x_2^{f_1} - M_{f_1}^1\right)^2 & \cdots & \left(x_n^{f_1} - M_{f_1}^1\right)^2 \\ \vdots & \vdots & \ddots & \vdots \\ \left(x_1^{f_{n_X}} - M_{f_{n_X}}^1\right)^2 & \left(x_2^{f_{n_X}} - M_{f_{n_X}}^1\right)^2 & \cdots & \left(x_{n_p}^{f_{n_X}} - M_{f_{n_X}}^1\right)^2 \\ \vdots & \vdots & \vdots & \vdots \\ \left(x_1^{f_1} - M_{f_1}^1\right)^k & \left(x_2^{f_1} - M_{f_1}^1\right)^k & \cdots & \left(x_{n_p}^{f_1} - M_{f_1}^1\right)^k \\ \vdots & \vdots & \ddots & \vdots \\ \left(x_1^{f_{n_X}} - M_{f_{n_X}}^1\right)^k & \left(x_2^{f_{n_X}} - M_{f_{n_X}}^1\right)^k & \cdots & \left(x_{n_p}^{f_{n_X}} - M_{f_{n_X}}^1\right)^k \end{bmatrix}_{(k \times n_X + 1) \times n_p}, C_{SLE} =$$

$$\begin{bmatrix} n_t \\ n_p M_{f_1}^1 \\ \vdots \\ n_p M_{f_{n_X}}^1 \\ (n_t - 1) M_{f_1}^2 \\ \vdots \\ (n_t - 1) M_{f_{n_X}}^2 \\ n_t \left(M_{f_1}^2\right)^{\frac{3}{2}} M_{f_1}^3 \\ \vdots \\ n_t \left(M_{f_{n_X}}^2\right)^{\frac{3}{2}} M_{f_{n_X}}^3 \\ n_t \left(M_{f_1}^2\right)^2 \left(M_{f_1}^4 + 3\right) \\ \vdots \\ n_t \left(M_{f_{n_X}}^2\right)^2 \left(M_{f_{n_X}}^4 + 3\right) \\ \vdots \\ n_t M_{f_{n_1}}^k \\ \vdots \\ n_t M_{f_{n_X}}^k \end{bmatrix}_{(k \times n_X + 1) \times 1}, P_{SLE} = \begin{bmatrix} p_{e_1} \\ \vdots \\ p_{e_{n_p}} \end{bmatrix}_{n_p \times 1}$$

In scenarios where the sample size is not predefined or where it is necessary to determine the maximum sample size, the sample size may be maximized instead of setting a fixed sample size. To address this, a linear programming (LP) problem, as detailed in Appendix 2, can be formulated and solved. This LP problem aims to minimize the negative $L1$ norm of the selection probabilities while satisfying equality constraints to identify the maximum size sample that fulfills the desired statistical moments for the selected features.

$$\begin{cases} \text{Minimize} & -\|P_{LP}\|^1 \\ \text{Such that:} & A_{LP} \cdot P_{LP} = C_{LP} \\ & \begin{bmatrix} 0 \\ \vdots \\ 0 \end{bmatrix}_{n_{\mathcal{P}} \times 1} \leq P_{LP} \leq \begin{bmatrix} 1 \\ \vdots \\ 1 \end{bmatrix}_{n_{\mathcal{P}} \times 1} \end{cases} \quad (2)$$

$$A_{LP} = \begin{bmatrix}
x_1^{f_1} - M_{f_1}^1 & x_2^{f_1} - M_{f_1}^1 & \cdots & x_{n_{\mathcal{P}}}^{f_1} - M_{f_1}^1 \\
\vdots & \vdots & \ddots & \vdots \\
x_1^{f_{n_X}} - M_{f_{n_X}}^1 & x_2^{f_{n_X}} - M_{f_{n_X}}^1 & \cdots & x_{n_{\mathcal{P}}}^{f_{n_X}} - M_{f_{n_X}}^1 \\
(x_1^{f_1} - M_{f_1}^1)^2 - M_{f_1}^2 & (x_2^{f_1} - M_{f_1}^1)^2 - M_{f_1}^2 & \cdots & (x_{n_{\mathcal{P}}}^{f_1} - M_{f_1}^1)^2 - M_{f_1}^2 \\
\vdots & \vdots & \ddots & \vdots \\
(x_1^{f_{n_X}} - M_{f_{n_X}}^1)^2 - M_{f_{n_X}}^2 & (x_2^{f_{n_X}} - M_{f_{n_X}}^1)^2 - M_{f_{n_X}}^2 & \cdots & (x_{n_{\mathcal{P}}}^{f_{n_X}} - M_{f_{n_X}}^1)^2 - M_{f_{n_X}}^2 \\
(x_1^{f_1} - M_{f_1}^1)^3 - (M_{f_1}^2)^{\frac{3}{2}} M_{f_1}^3 & (x_2^{f_1} - M_{f_1}^1)^3 - (M_{f_1}^2)^{\frac{3}{2}} M_{f_1}^3 & \cdots & (x_{n_{\mathcal{P}}}^{f_1} - M_{f_1}^1)^3 - (M_{f_1}^2)^{\frac{3}{2}} M_{f_1}^3 \\
\vdots & \vdots & \ddots & \vdots \\
(x_1^{f_{n_X}} - M_{f_{n_X}}^1)^3 - (M_{f_{n_X}}^2)^{\frac{3}{2}} M_{f_{n_X}}^3 & (x_2^{f_{n_X}} - M_{f_{n_X}}^1)^3 - (M_{f_{n_X}}^2)^{\frac{3}{2}} M_{f_{n_X}}^3 & \cdots & (x_{n_{\mathcal{P}}}^{f_{n_X}} - M_{f_{n_X}}^1)^3 - (M_{f_{n_X}}^2)^{\frac{3}{2}} M_{f_{n_X}}^3 \\
(x_1^{f_1} - M_{f_1}^1)^4 - (M_{f_1}^2)^2 (M_{f_1}^4 + 3) & (x_2^{f_1} - M_{f_1}^1)^4 - (M_{f_1}^2)^2 (M_{f_1}^4 + 3) & \cdots & (x_{n_{\mathcal{P}}}^{f_1} - M_{f_1}^1)^4 - (M_{f_1}^2)^2 (M_{f_1}^4 + 3) \\
\vdots & \vdots & \ddots & \vdots \\
(x_1^{f_{n_X}} - M_{f_{n_X}}^1)^4 - (M_{f_{n_X}}^2)^2 (M_{f_{n_X}}^4 + 3) & (x_2^{f_{n_X}} - M_{f_{n_X}}^1)^4 - (M_{f_{n_X}}^2)^2 (M_{f_{n_X}}^4 + 3) & \cdots & (x_{n_{\mathcal{P}}}^{f_{n_X}} - M_{f_{n_X}}^1)^4 - (M_{f_{n_X}}^2)^2 (M_{f_{n_X}}^4 + 3) \\
(x_1^{f_1} - M_{f_1}^1)^k - M_{f_1}^k & (x_2^{f_1} - M_{f_1}^1)^k - M_{f_1}^k & \cdots & (x_{n_{\mathcal{P}}}^{f_1} - M_{f_1}^1)^k - M_{f_1}^k \\
\vdots & \vdots & \ddots & \vdots \\
(x_1^{f_{n_X}} - M_{f_{n_X}}^1)^k - M_{f_{n_X}}^k & (x_2^{f_{n_X}} - M_{f_{n_X}}^1)^k - M_{f_{n_X}}^k & \cdots & (x_{n_{\mathcal{P}}}^{f_{n_X}} - M_{f_{n_X}}^1)^k - M_{f_{n_X}}^k
\end{bmatrix}_{(k \times n_X) \times n_{\mathcal{P}}}$$

$$C_{LP} = \begin{bmatrix} 0 \\ \vdots \\ 0 \\ -M_{f_1}^2 \\ \vdots \\ -M_{f_{n_X}}^2 \\ 0 \\ \vdots \\ 0 \end{bmatrix}_{(k \times n_X) \times 1}, \quad P_{LP} = \begin{bmatrix} p_{e_1} \\ \vdots \\ p_{e_{n_{\mathcal{P}}}} \end{bmatrix}_{n_{\mathcal{P}} \times 1}$$

However, depending on the application, strict equality may not be necessary, especially since statistical equivalence does not demand it. Therefore, the above equality-constrained LP problem can be relaxed to an inequality-constrained LP problem, as discussed in Appendix 2.

$$\begin{cases} Minimize \ -(\|P_{LP}\|^1 - \beta \|\eta\|^1) & (3) \\ Such \ that: \ A_{LP} \cdot P_{LP} - C_{LP} \leq \eta \\ \qquad \qquad A_{LP} \cdot P_{LP} - C_{LP} \geq -\eta \\ \begin{bmatrix} 0 \\ \ldots \\ 0 \end{bmatrix}_{n_p \times 1} \leq P_{LP} \leq \begin{bmatrix} 1 \\ \ldots \\ 1 \end{bmatrix}_{n_p \times 1} \\ \begin{bmatrix} 0 \\ \ldots \\ 0 \end{bmatrix}_{(k \times n_X) \times 1} \leq \eta \leq \eta_{max} = \begin{bmatrix} \eta_{max_1} \\ \ldots \\ \eta_{max_m} \end{bmatrix}_{(k \times n_X) \times 1} \end{cases}$$

Where $\eta$ is the new decision variable that defines the permissible error in features of interest, and $\eta_{max}$ representing the maximum allowable error. The parameter $\beta$ is a penalization term between $P_{LP}$ and $\eta$, that trades off between population size and error. This trade-off can be fine-tuned depending on different applications. The flowchart of the DSPS is presented in Appendix 3.

In theory, the same LP approach can be used to minimize the sample size. However, in all the equations, the term $E\left(\sum_{i=1}^{n_p} b_{e_i}\right) = \sum_{i=1}^{n_p} p_{e_i}$ is used to approximate the expected sample size with the summation selection probabilities. According to the law of large numbers (LLN), this approximation is accurate if the sample size is sufficiently large — the more samples we have, the more accurate this approximation becomes[31]. Regarding this, while minimizing the sample size, if the sum of probabilities remains sufficiently large (usually more than 30), this can be considered as the minimum sample size. However, if the sum becomes too small, the $E\left(\sum_{i=1}^{n_p} b_{e_i}\right) = \sum_{i=1}^{n_p} p_{e_i}$ does not hold anymore. As a result, the probabilities may satisfy the constraints, but the distribution of the features may deviate from the target distribution.

## 2.2. Clinical trial data and Protocol

The clinical trial data utilized in this paper originates from a randomized, 26-week, treat-to-target trial (NCT01336023) conducted on insulin-naïve patients with T2D [32]. The trial evaluated the efficacy and safety of a fixed-ratio combination of insulin Degludec and the GLP-1 receptor agonist Liraglutide (IDegLira) compared to each component alone, across three treatment arms. As part of the trial protocol, insulin Degludec (IDeg) in the IDeg arm was titrated twice weekly to achieve a fasting blood glucose (FBG) target of 72-90 mg/dl, based on the titration rule reported in Table 1.

The data reported in the publication of this trial encompasses both demographic and glycemic behaviors. Demographic data includes variables such as age, body weight, body mass index (BMI), and the duration of T2D. Additionally, Glycemic behavior data focuses on key metrics such as HbA1c levels, Fasting Blood Glucose (FBG) readings, the incidence of hypoglycemic events (FBG < 3.1 mmol/L), and the insulin dosages administered throughout the trial. These metrics were carefully documented at baseline and at regular intervals during the trial, specifically at weeks 4, 8, 12, 16, 20, and 26. Notably, the publication reports the mean and standard deviation for all measures except for hypoglycemic events.

Table 11. PROTOCOL-SPECIFIED TITRATION ALGORITHM FOR INSULIN DEGLUDEC AS USED IN CLINICAL TRIAL NCT01336023.

| Measurement | Threshold (mg/dl) | Dose change (unit) |
| --- | --- | --- |
| Any pre-breakfast SMBG | $< 72$ | $-2$ |
| Mean pre-breakfast SMBG | $\geq 72\ \&\ \leq 90$ | $0$ |
| Mean pre-breakfast SMBG | $> 72$ | $+2$ |

### 2.3. UVLab and Simulation Setup

The UVA Virtual Lab (UVLab) is a newly developed simulation platform designed to replicate the metabolic behavior of individuals with T2D. This platform is based on mathematical models that describe the fluxes, clearance of glucose, insulin, and other hormones influencing these processes [33]. These mathematical models depend on a set of parameters that characterize different metabolic behaviors, and each set of parameters defines a virtual subject (avatar). This platform is equipped with 6062 virtual subjects (avatars) accounting for heterogeneity and different phenotypes of T2D.

The simulation setup consists of personalized meals, an insulin dosing regimen, and day-to-day variability to accurately reflect real-world conditions. In this experiment, each virtual subject consumed a personalized meal plan corresponding to their baseline HbA1c levels, consisting of three main meals and one snack, with carbohydrate distribution ratios of 0.3, 0.3, 0.3, and 0.1, respectively. virtual subject measured their FBG before their first meal of the day and administered insulin Degludec as their daily basal insulin. The Endogenous Glucose Production (EGP) steady state was perturbed using a normal distribution centered around the nominal average value for each virtual subject, with a coefficient of variation (CV) of 10%, to ensure achieving an average CV of 14% for fasting plasma glucose (FPG), which was observed in real data.

Simulation outcomes included: fasting Plasma Glucose (FPG) measured using a simulated blood glucose meter with noise settings calibrated according to those reported in [34], ensuring realistic variability in the measurements. Insulin refers to the total administered daily insulin dose, which includes the personalized basal insulin regimen for each virtual subject. The Glycemic Management Indicator (GMI) was calculated using the formula $GMI\% = 3.31 + 0.02392 \times [mean\ glucose\ in\ mg/dL]$, providing an estimate of HbA1c [33]. Hypoglycemia events were defined as instances where FPG levels fell below 3.1 mmol/L, reflecting clinically significant low blood sugar episodes. To remove the effect of the randomness in the simulation's outcomes, all the simulations were repeated 10 times with different random seeds, and the metrics were averaged across the 10 runs to ensure robustness and accuracy.

### 2.4. Strategies in Applying DSPS

Two strategies are applied within the DSPS algorithm. The first approach uses all available data from the trial, referred to as reproduction. The second approach uses only the data from the initial weeks of the trial to select a sub-population, allowing for testing the predictive capacity of the simulation platform by forecasting the trial's outcomes over the remaining duration. The LP problem 3 is utilized in both strategies, with the hyperparameters set as in Table 2. The hyperparameter $\alpha$ was set to 5% of the clinical trial sample size. This choice was made to ensure that, if the maximum population size selected by the DSPS is equal to the actual clinical trial size, the error between the sub-population mean and variance and the target moments remains below 5%.

Table 2. Hyper parameters used in the experiment setup.

| Symbol | Description | Value |
|---|---|---|
| $\beta$ | Tradeoff error-population size | $\left[\dfrac{1}{M_{f_1}^1 + \varepsilon} \quad \cdots \quad \dfrac{1}{M_{f_{n_X}}^1 + \varepsilon} \quad \dfrac{1}{M_{f_1}^2 + \varepsilon} \quad \cdots \quad \dfrac{1}{M_{f_{n_X}}^2 + \varepsilon}\right]$ |
| $\alpha$ | Tradeoff error-population size | $5\% \times Clinical\ Trial\ Sample\ size$ |
| $\eta_{max}$ | Maximum tolerated error | $\alpha \times \left[M_{f_1}^1 + \varepsilon \quad \cdots \quad M_{f_{n_X}}^1 + \varepsilon \quad M_{f_1}^2 + \varepsilon \quad \cdots \quad M_{f_{n_X}}^2 + \varepsilon\right]$ |
| $\varepsilon$ | Ensure non-zero denominator | $10^{-6}$ |

### 2.4.1. Reproduction of the Clinical Trial

In the reproduction strategy, the initial goal is to match all available metrics at baseline and throughout the trial's duration to faithfully replicate the clinical trial within the UVLab. These metrics include demographic data such as age, body weight, BMI, and the duration of T2D, both in terms of mean and variance. Additionally, glycemic control metrics are considered, which encompass HbA1c levels, Fasting Blood Glucose (FBG), the cumulative number of hypoglycemic events per patient (defined as FBG < 3.1 mmol/L), and the daily units of insulin Degludec administered at specific weeks (0, 4, 8, 12, 16, 20, 26).

### 2.4.2. Prediction of the Clinical Trial

The approach begins by using data from the initial weeks of the trial to match the relevant glycemic metrics within the UVLab. The selected sub-population is identified based on available data up to a certain week, ensuring that it closely aligns with the observed clinical trial outcomes during that period. Once this initial alignment is achieved, the sub-population is then used to predict the outcomes for the remaining weeks of the trial.

### 2.5. Outcome Metrics

The outcome metrics are designed to evaluate how closely the selected sub-population metrics align with those used in the selection process. The primary metrics of closeness employed are the Relative Sum of Square Errors (RSSE) and Percentage Error.

**Relative Sum of Square Errors (RSSE):**

RSSE quantifies the distance between the moments of the features in the selected sub-population and the target moments. It is calculated using the following formula:

$$RSSE = \sum_{j=1}^{n_X} \sum_{i=1}^{k_j} \left( \frac{\left(M_{f_j,sample}^{k_i} - M_{f_j}^{k_i}\right)^2}{\left(M_{f_j}^{k_i}\right)^2} \right)$$

Where $M_{f_j,sample}^{k_i}$ represents $k_i$-th moment of the $f_j$ feature in the selected sub-population.

**Percentage Error:**

The Percentage Error expresses the deviation between the moments of the features in the selected sub-population and the target moments as a percentage. It is calculated using the following formula:

$$PE_{f_j}^{k_i} = \frac{\left|M_{f_j,sample}^{k_i} - M_{f_j}^{k_i}\right|}{M_{f_j}^{k_i}} \times 100\%$$

Where $PE_{f_j}^{k_i}$ represents the percentage error between the $k_i$-th moment of the $f_j$ feature in the selected sub-population and its corresponding target moment.

Both RSSE and Percentage Error are also used to evaluate the predictability of the selected sub-population. By applying these metrics, it is possible to assess how well the selected group can forecast the clinical trial outcomes based on the early-stage data, further validating the effectiveness of the prediction strategy within the DSPS framework.

## 3. Results

The results of the DSPS method, applied both for reproducing and predicting the outcomes of the clinical trial, are presented below.

### 3.1. Reproduction of the Clinical Trial

In this section, the outcomes of the population selection process using DSPS to replicate all available outcomes of the clinical trial are presented. The results are summarized in Figure 1. The Relative Sum of Square Errors (RSSE) and Percentage Error for the selected population are 0.33 and 1.07% (1.2%) respectively, in terms of mean and standard deviation. These values represent the differences between the features used in the selection process and the corresponding features in the selected sub-population.

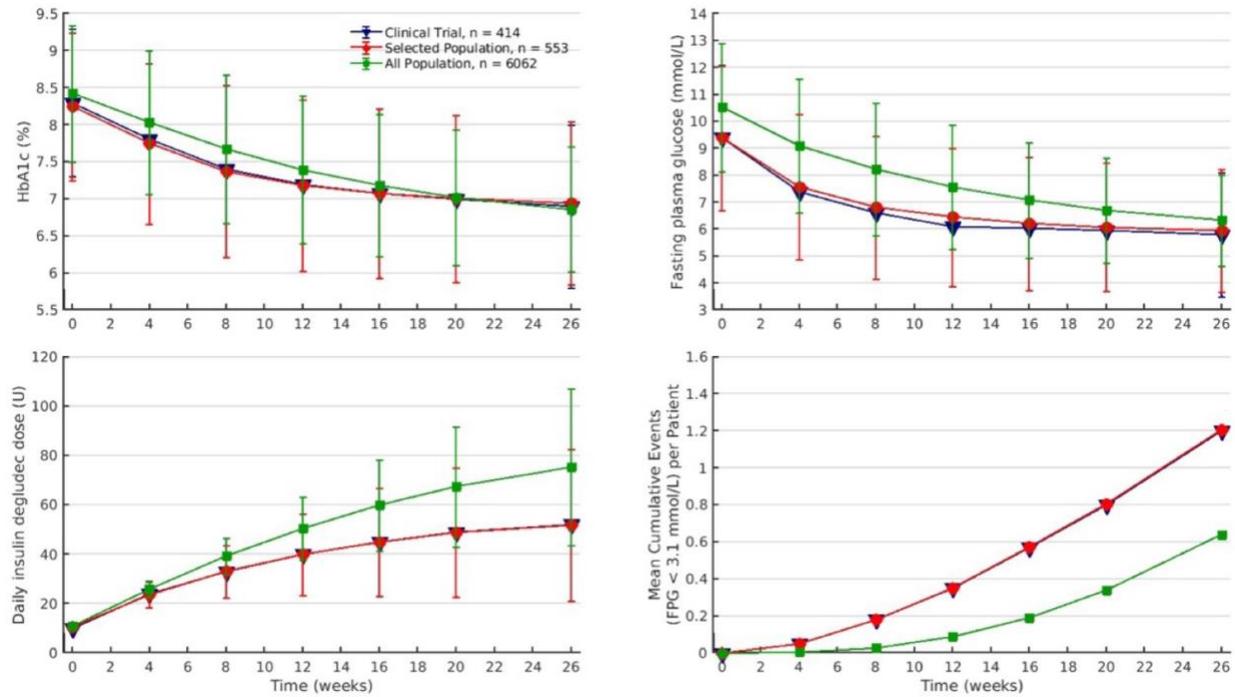

**Figure 1. Glycemic efficacy, insulin doses, and cumulative hypoglycemia over 26 weeks in Clinical trial, Selected population, and All population at UVlab**

### 3.2. Prediction of the Clinical Trial

In this section, the DSPS was employed to select a population using the data in the initial weeks and then predict the outcomes of the rest of the trial. In Figure 2, the summary of sub-population selection across different weeks of data used to select is described. In Figure 3, an example of prediction by only using the week 0 and 4 data is presented. The RSSE and Percentage Error in this context are calculated based on the differences between the predicted features and the actual features observed in the later weeks of the clinical trial.

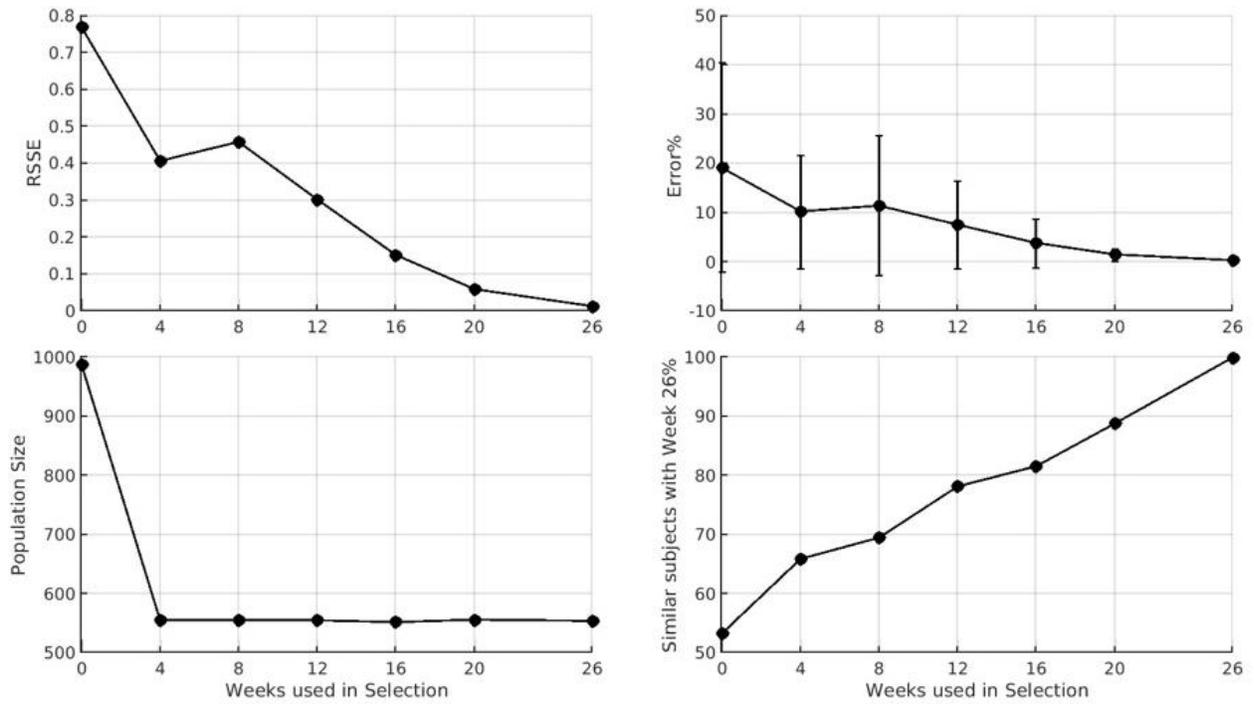

**Figure 2. Summary of selected population and the prediction error**

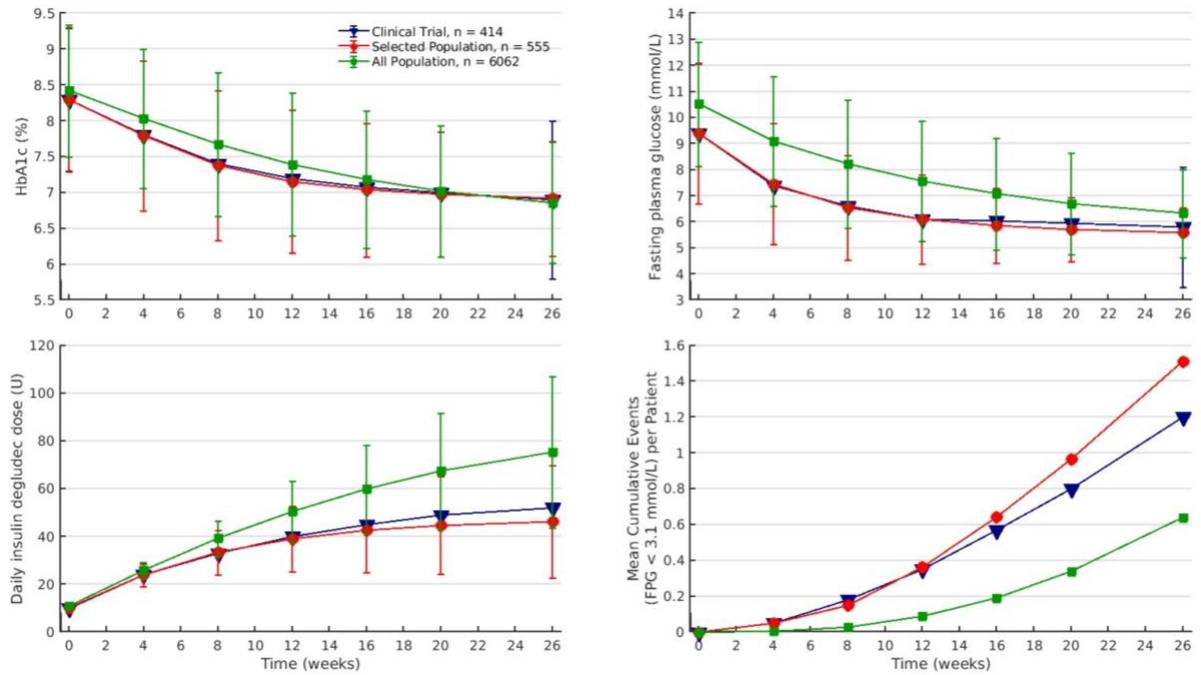

**Figure 3. Glycemic efficacy, insulin doses, and cumulative hypoglycemia over 26 weeks in Clinical trial, Selected population, and All population at UVLab**

## 4. Discussion

This paper introduced and validated the Distribution-Based Sub-Population Selection (DSPS) method as a tool for selecting samples from a parent population with specific features. This tool is specifically applied in this work to reproduce and predict the outcomes of a clinical trial using the UVLab, which provides a parent population of 6,062 virtual subjects. The application of the DSPS method successfully reproduced the clinical trial's outcomes with a Relative Sum of Square Errors (RSSE) of 0.33 and a Percentage Error of 1.07% (1.2%) in terms of mean and standard deviation. Additionally, when used for prediction, the method demonstrated its capability to select a sub-population that can forecast trial outcomes based on initial weeks of data.

Sub-population selection within simulation platforms holds great potential for advancing personalized medicine in the field of T2D. Using statistical methods to identify virtual populations with similar clinical behavior as observed in real clinical trials, these platforms can enable personalized interventions for the selected populations. For example, virtual populations can be used to identify patient subgroups that may respond differently to certain treatments based on their baseline characteristics and clinical behaviors. This information can guide the development of tailored interventions that lead to improved glycemic control. Additionally, virtual populations can facilitate efficient and controlled experimentation, allowing for the rapid assessment of different treatment strategies in a cost-effective and ethical manner with higher statistical power [34]. This can accelerate the development of new treatment approaches and provide valuable insights into the effectiveness of different interventions across T2D subgroups.

In this work, two different applications, reproduction and prediction, were employed for the use of DSPS. The rationale behind the reproduction strategy lies in its aim to find a sub-population that closely matches all aspects of the clinical trial data, effectively replicating the trial within the simulation platform. This comprehensive matching process evaluates the UVLab's ability to replicate clinical trial data, incorporating all available information to verify the existence of a sub-population within the UVLab that mirrors the complete dataset of a clinical trial at the population level. Once this population is identified, it can serve as a foundation for further analyses related to that specific trial, such as testing different titration algorithms or insulin types.

Moreover, the rationale behind focusing on the prediction of clinical trial outcomes lies in its potential to streamline the trial design process. Predicting the outcome of a clinical trial from early data can offer significant insights into the effectiveness of the intervention, potentially reducing the duration and cost of trials. This capability is particularly useful for adapting ongoing trials, optimizing trial parameters, or even deciding on early termination if results are not promising. By accurately forecasting outcomes, researchers can make informed decisions that enhance the efficiency of the clinical trial process, thereby accelerating the development of new treatments.

In addressing the problem of sub-population selection, one could approach it as an Integer Programming (IP) problem, where decision variables represent the presence or absence of each virtual subject. However, ILP is an NP-hard problem [35], meaning it is computationally intensive to solve efficiently, especially when dealing with large populations, such as the 6,062 virtual subjects in the UVLab. Introducing softness in the decision variables by allowing probabilities instead of binary decisions transforms the problem into a Linear Programming (LP) problem, which is computationally feasible (P class problem).

Linear programming was chosen for this work over quadratic programming (QP) for several key reasons. First, LP is computationally less expensive, crucial when dealing with a large number of decision variables. Second, LP inherently promotes sparsity in the solution, with selection probabilities tending toward 0 or 1. This sparsity can be advantageous, as it makes the selected samples more robust and reduces the effects of random number generation, bringing the problem closer to an ILP form without the computational challenges of NP-hardness. However, the downside is that in real-world situations, probabilities are often not as sparse, and it may be desirable to give a reasonable chance of selection to almost every subject.

One limitation of this work stems from the inherent constraints of the simulation platforms, which often lack several features that exist in real-world scenarios. Firstly, many physiological components are either not modeled or are approximated within the simulator. For example, the model of day-to-day variability used in this work, while effective, requires further validation to ensure it accurately reflects real-world physiological fluctuations. Additionally, behavioral aspects are largely oversimplified or omitted entirely. In this study, we assumed 100% adherence to the intervention protocols, such as perfect compliance with insulin titration rules. However, in reality, patient adherence is far from perfect, and such assumptions may not hold true, potentially limiting the simulator's ability to fully capture real-world outcomes. These limitations highlight the need for ongoing development and validation of simulation platforms to incorporate more complex and realistic physiological and behavioral factors.

In conclusion, this paper presents and validates the Distribution-Based Sub-Population Selection (DSPS) method as a robust tool for selecting samples from a parent population to reproduce and predict clinical trial outcomes. By applying this method within the UVLab simulation platform, we successfully demonstrated its capability to replicate the results of a clinical trial and predict future outcomes based on early data. The DSPS method's effectiveness in both reproducing trial data and forecasting results underscores its potential as a valuable tool in the field of personalized medicine, particularly for optimizing treatment strategies and designing more efficient clinical trials. To promote further research and application, the implemented version of the DSPS algorithm has been released on [GitHub](GitHub).

**Acknowledgements:**

This work was funded by a grant from Novo Nordisk.

**Appendix 1:**

Assuming a population of size $n_t$, with similar moments for the distribution of the variables of interest is desired. To achieve this, we can formulate a set of equations that must be satisfied by the $p_{e_i}$ values.

$$n_s = E\left(\sum_{i=1}^{n_p} b_{e_1}\right) = \sum_{i=1}^{n_p} E(b_{e_1}) = \sum_{i=1}^{n_p} p_{e_1} + 0 \times (1 - p_{e_1}) = \sum_{i=1}^{n_p} p_{e_1} = n_t \qquad (4)$$

$$M_{f_j,s}^1 = E\left(\sum_{i=1}^{n_p} \frac{p_{e_i} \cdot x_i^{f_j}}{n_t}\right) = \frac{1}{n_t}\sum_{i=1}^{n_p} x_i^{f_j} E(b_{e_i}) = \frac{1}{n_t}\sum_{i=1}^{n_p} x_i^{f_j} \cdot p_{e_i} = M_{f_j}^1 \quad (5)$$

$$M_{f_j,s}^2 = E\left(\sum_{i=1}^{n_p} \frac{p_{e_i} \cdot \left(x_i^{f_j} - M_{f_j,s}^1\right)^2}{n_t - 1}\right) = \frac{1}{n_t - 1}\sum_{i=1}^{n_p} p_{e_i} \cdot \left(x_i^{f_j} - M_{f_j,s}^1\right)^2 = M_{f_j}^2 \quad (6)$$

$$M_{f_j,s}^3 = E\left(\sum_{i=1}^{n_p} \frac{p_{e_i} \cdot \left(x_i^{f_j} - M_{f_j,s}^1\right)^3}{n_t}\right) = \frac{1}{n_t}\sum_{i=1}^{n_p} p_{e_i} \cdot \left(x_i^{f_j} - M_{f_j,s}^1\right)^3 = M_{f_j}^3 \quad (7)$$

$$M_{f_j,s}^4 = E\left(\sum_{i=1}^{n_p} \frac{p_{e_i} \cdot \left(x_i^{f_j} - M_{f_j,s}^1\right)^4}{n_t \times \left(M_{f_j}^2\right)^2}\right) - 3 = \frac{1}{n_t \times \left(M_{f_j}^2\right)^2}\sum_{i=1}^{nn_p} p_{e_i} \cdot \left(x_i^{f_j} - M_{f_j,s}^1\right)^4 - 3 = M_{f_j}^4 \quad (8)$$

$$M_{f_j,s}^5 = E\left(\sum_{i=1}^{n_p} \frac{p_{e_i} \cdot \left(x_i^{f_j} - M_{f_j,s}^1\right)^5}{n_t}\right) = \frac{1}{n_t}\sum_{i=1}^{n_p} p_{e_i} \cdot \left(x_i^{f_j} - M_{f_j,s}^1\right)^5 = M_{f_j}^5 \quad (9)$$

The above equations can be written in the system of linear equations 1.

**Appendix 2:**

In order to relax the hard equality constraint to the soft inequality constraint, the following steps are done.

$$\begin{cases} Minimize \quad -\|P_{LP}\|^1 \\ Such\ that: A_{LP} \cdot P_{LP} = C_{LP} \\ \begin{bmatrix} 0 \\ \vdots \\ 0 \end{bmatrix}_{n_p \times 1} \leq P_{LP} \leq \begin{bmatrix} 1 \\ \vdots \\ 1 \end{bmatrix}_{n_p \times 1} \end{cases} \quad (10)$$

$$\begin{cases} Minimize \quad \|-P_{LP}\|^1 \\ Such\ that: A_{LP} \cdot P_{LP} - C_{LP} \leq \eta \\ \qquad\qquad A_{LP} \cdot P_{LP} - C_{LP} \geq -\eta \\ P_{LP} \leq \begin{bmatrix} 1 \\ \vdots \\ 1 \end{bmatrix}_{n_p \times 1} \\ P_{LP} \geq \begin{bmatrix} 0 \\ \vdots \\ 0 \end{bmatrix}_{n_p \times 1} \\ \eta \leq \eta_{max} = \begin{bmatrix} \eta_{max_1} \\ \ldots \\ \eta_{max_m} \end{bmatrix}_{(k \times n_X) \times 1} \\ \eta \geq \begin{bmatrix} 0 \\ \vdots \\ 0 \end{bmatrix}_{(k \times n_X) \times 1} \end{cases} \quad (11)$$

$$\begin{cases} \text{Minimize } -(\|P_{LP}\|^1 - \beta\|\eta\|^1) \\ \text{Such that: } A_{LP} \cdot P_{LP} - C_{LP} \leq \eta \\ \phantom{\text{Such that: }} A_{LP} \cdot P_{LP} - C_{LP} \geq -\eta \\ \begin{bmatrix} 0 \\ \vdots \\ 0 \end{bmatrix}_{n_p \times 1} \leq P_{LP} \leq \begin{bmatrix} 1 \\ \vdots \\ 1 \end{bmatrix}_{n_p \times 1} \\ \begin{bmatrix} 0 \\ \vdots \\ 0 \end{bmatrix}_{(k \times n_X) \times 1} \leq \eta \leq \eta_{max} = \begin{bmatrix} \eta_{max_1} \\ \vdots \\ \eta_{max_m} \end{bmatrix}_{(k \times n_X) \times 1} \end{cases} \quad (12)$$

**Appendix 3:**

This flowchart outlines the process for selecting a sample from a parent distribution, with variations depending on whether sample size is predefined or not, while also considering specified hyperparameters for optimization

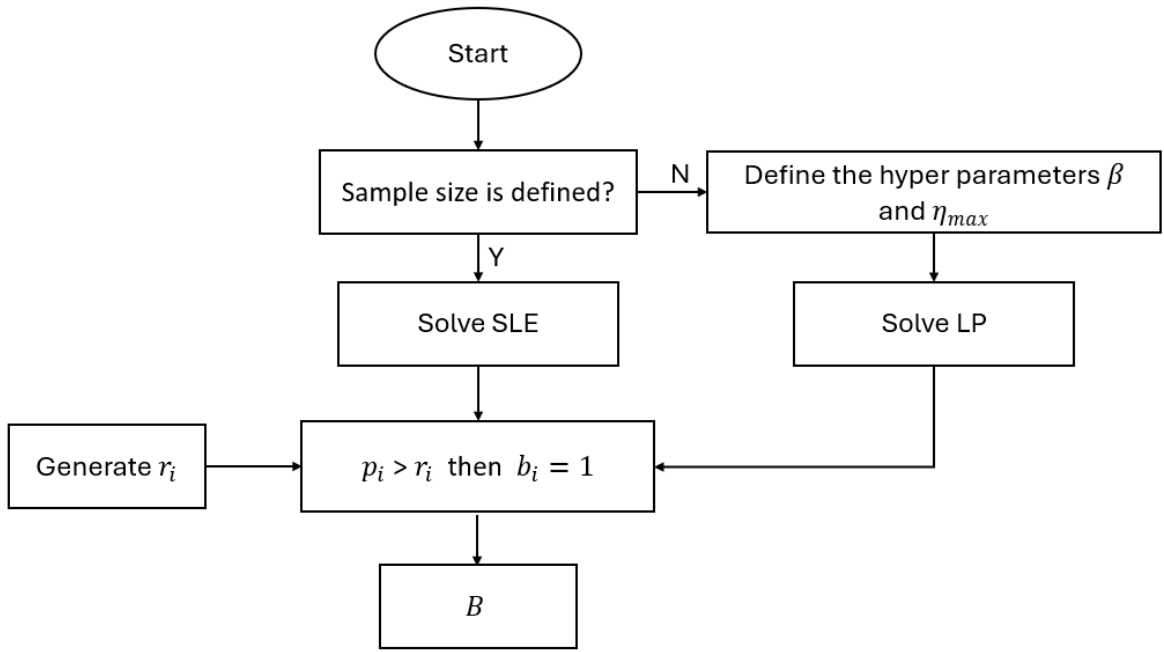

# References

[1] E. P. Joslin, "The Treatment of Diabetes Mellitus," *Can. Med. Assoc. J.*, vol. 6, no. 8, pp. 673–684, Aug. 1916, Accessed: Aug. 23, 2024. [Online]. Available: https://www.ncbi.nlm.nih.gov/pmc/articles/PMC1584654/

[2] R. A. DeFronzo et al., "Type 2 diabetes mellitus," *Nat. Rev. Dis. Primer*, vol. 1, p. 15019, Jul. 2015, doi: 10.1038/nrdp.2015.19.

[3] "Diabetes Update: Prevention and Management of Diabetes Complications." Accessed: Aug. 23, 2024. [Online]. Available: https://www.researchgate.net/publication/318862048_Diabetes_Update_Prevention_and_Management_of_Diabetes_Complications

[4] I. Vargatu, "WILLIAMS TEXTBOOK OF ENDOCRINOLOGY," *Acta Endocrinol. Buchar.*, vol. 12, no. 1, p. 113, 2016, doi: 10.4183/aeb.2016.113.

[5] "Intensive blood-glucose control with sulphonylureas or insulin compared with conventional treatment and risk of complications in patients with type 2 diabetes (UKPDS 33). UK Prospective Diabetes Study (UKPDS) Group," *Lancet Lond. Engl.*, vol. 352, no. 9131, pp. 837–853, Sep. 1998.

[6] Diabetes Control and Complications Trial Research Group et al., "The effect of intensive treatment of diabetes on the development and progression of long-term complications in insulin-dependent diabetes mellitus," *N. Engl. J. Med.*, vol. 329, no. 14, pp. 977–986, Sep. 1993, doi: 10.1056/NEJM199309303291401.

[7] J. M. Lachin, S. Genuth, D. M. Nathan, B. Zinman, B. N. Rutledge, and DCCT/EDIC Research Group, "Effect of glycemic exposure on the risk of microvascular complications in the diabetes control and complications trial--revisited," *Diabetes*, vol. 57, no. 4, pp. 995–1001, Apr. 2008, doi: 10.2337/db07-1618.

[8] Y. Chu, S. Li, J. Tang, and H. Wu, "The potential of the Medical Digital Twin in diabetes management: a review," *Front. Med.*, vol. 10, p. 1178912, Jul. 2023, doi: 10.3389/fmed.2023.1178912.

[9] A. Vallée, "Digital twin for healthcare systems," *Front. Digit. Health*, vol. 5, p. 1253050, Sep. 2023, doi: 10.3389/fdgth.2023.1253050.

[10] B. Kovatchev, "A Century of Diabetes Technology: Signals, Models, and Artificial Pancreas Control," *Trends Endocrinol. Metab.*, vol. 30, no. 7, pp. 432–444, Jul. 2019, doi: 10.1016/j.tem.2019.04.008.

[11] "Validation of the UVA Simulation Replay Methodology Using Clinical Data: Reproducing a Randomized Clinical Trial | Diabetes Technology & Therapeutics." Accessed: Aug. 23, 2024. [Online]. Available: https://www.liebertpub.com/doi/10.1089/dia.2023.0595

[12] "ReplayBG: A Digital Twin-Based Methodology to Identify a Personalized Model From Type 1 Diabetes Data and Simulate Glucose Concentrations to Assess Alternative Therapies | IEEE Journals & Magazine | IEEE Xplore." Accessed: Aug. 23, 2024. [Online]. Available: https://ieeexplore.ieee.org/document/10164140

[13] "Takagi-Sugeno (TS) Fuzzy model-basedobserver design for glucose-insulin system in diabetes type 1: An LMI approach | IEEE Conference Publication | IEEE Xplore." Accessed: Aug. 23, 2024. [Online]. Available: https://ieeexplore.ieee.org/document/9260768

[14] M. W. Percival et al., "Development of a multi-parametric model predictive control algorithm for insulin delivery in type 1 diabetes mellitus using clinical parameters," *J. Process Control*, vol. 21, no. 3, pp. 391–404, Mar. 2011, doi: 10.1016/j.jprocont.2010.10.003.

[15] K. Turksoy, L. Quinn, E. Littlejohn, and A. Cinar, "Multivariable adaptive identification and control for artificial pancreas systems," *IEEE Trans. Biomed. Eng.*, vol. 61, no. 3, pp. 883–891, Mar. 2014, doi: 10.1109/TBME.2013.2291777.

[16] Q. Wang *et al.*, "Personalized State-space Modeling of Glucose Dynamics for Type 1 Diabetes Using Continuously Monitored Glucose, Insulin Dose, and Meal Intake," *J. Diabetes Sci. Technol.*, vol. 8, no. 2, pp. 331–345, Mar. 2014, doi: 10.1177/1932296814524080.

[17] A. El Fathi, M. Raef Smaoui, V. Gingras, B. Boulet, and A. Haidar, "The Artificial Pancreas and Meal Control: An Overview of Postprandial Glucose Regulation in Type 1 Diabetes," *IEEE Control Syst. Mag.*, vol. 38, no. 1, pp. 67–85, Feb. 2018, doi: 10.1109/MCS.2017.2766323.

[18] A. E. Fathi, R. E. Kearney, E. Palisaitis, B. Boulet, and A. Haidar, "A Model-Based Insulin Dose Optimization Algorithm for People With Type 1 Diabetes on Multiple Daily Injections Therapy," *IEEE Trans. Biomed. Eng.*, vol. 68, no. 4, pp. 1208–1219, Apr. 2021, doi: 10.1109/TBME.2020.3023555.

[19] M. Ganji and M. Pourgholi, "An LMI-based Robust Fuzzy Controller for Blood Glucose Regulation in Type 1 Diabetes," Aug. 19, 2024, *arXiv*: arXiv:2408.10333. doi: 10.48550/arXiv.2408.10333.

[20] "The UVA/PADOVA Type 1 Diabetes Simulator: New Features - PubMed." Accessed: Aug. 23, 2024. [Online]. Available: https://pubmed.ncbi.nlm.nih.gov/24876534/

[21] R. Visentin *et al.*, "The UVA/Padova Type 1 Diabetes Simulator Goes From Single Meal to Single Day," *J. Diabetes Sci. Technol.*, vol. 12, no. 2, pp. 273–281, Mar. 2018, doi: 10.1177/1932296818757747.

[22] R. Visentin, M. Schiavon, C. Giegerich, T. Klabunde, C. D. Man, and C. Cobelli, "Long-acting Insulin in Diabetes Therapy: In Silico Clinical Trials with the UVA/Padova Type 1 Diabetes Simulator," *Annu. Int. Conf. IEEE Eng. Med. Biol. Soc. IEEE Eng. Med. Biol. Soc. Annu. Int. Conf.*, vol. 2018, pp. 4905–4908, Jul. 2018, doi: 10.1109/EMBC.2018.8513234.

[23] R. Visentin, M. Schiavon, C. Giegerich, T. Klabunde, C. D. Man, and C. Cobelli, "Incorporating Long-Acting Insulin Glargine Into the UVA/Padova Type 1 Diabetes Simulator for In Silico Testing of MDI Therapies," *IEEE Trans. Biomed. Eng.*, vol. 66, no. 10, pp. 2889–2896, Oct. 2019, doi: 10.1109/TBME.2019.2897851.

[24] M. D. Breton, R. Hinzmann, E. Campos-Nañez, S. Riddle, M. Schoemaker, and G. Schmelzeisen-Redeker, "Analysis of the Accuracy and Performance of a Continuous Glucose Monitoring Sensor Prototype: An In-Silico Study Using the UVA/PADOVA Type 1 Diabetes Simulator," *J. Diabetes Sci. Technol.*, vol. 11, no. 3, pp. 545–552, May 2017, doi: 10.1177/1932296816680633.

[25] C. Dalla Man, R. A. Rizza, and C. Cobelli, "Meal simulation model of the glucose-insulin system," *IEEE Trans. Biomed. Eng.*, vol. 54, no. 10, pp. 1740–1749, Oct. 2007, doi: 10.1109/TBME.2007.893506.

[26] R. Visentin, C. Cobelli, and C. Dalla Man, "The Padova Type 2 Diabetes Simulator from Triple-Tracer Single-Meal Studies: In Silico Trials Also Possible in Rare but Not-So-Rare Individuals," *Diabetes Technol. Ther.*, vol. 22, no. 12, pp. 892–903, Dec. 2020, doi: 10.1089/dia.2020.0110.

[27] "Partitioning glucose distribution/transport, disposal, and endogenous production during IVGTT | American Journal of Physiology-Endocrinology and Metabolism." Accessed: Aug. 23, 2024. [Online]. Available: https://journals.physiology.org/doi/full/10.1152/ajpendo.00304.2001

[28] A. N. Shahidehpour, "A Type 2 Diabetes Simulator for Multiscale Glycemic Dynamics," Ph.D., Illinois Institute of Technology, United States -- Illinois, 2024. Accessed: Aug. 23, 2024. [Online]. Available: https://www.proquest.com/docview/3092163779/abstract/8644EA3E3DAC418APQ/1

[30] M. GANJIARJENAKI, C. FABRIS, A. EL FATHI, D. LV, B. KOVATCHEV, and M. D. BRETON, "75-OR: In-Silico Replay of Insulin Degludec and Liraglutide Clinical Trial Data in Subjects with Type 2 Diabetes," *Diabetes*, vol. 73, no. Supplement_1, 2024, Accessed: Aug. 23, 2024. [Online]. Available: https://diabetesjournals.org/diabetes/article/73/Supplement_1/75-OR/156368

[31] P. Sedlmeier and G. Gigerenzer, "Intuitions about sample size: the empirical law of large numbers," *J. Behav. Decis. Mak.*, vol. 10, no. 1, pp. 33–51, 1997, doi: 10.1002/(SICI)1099-0771(199703)10:1<33::AID-BDM244>3.0.CO;2-6.

[32] S. C. L. Gough *et al.*, "Efficacy and safety of a fixed-ratio combination of insulin degludec and liraglutide (IDegLira) compared with its components given alone: results of a phase 3, open-label,

randomised, 26-week, treat-to-target trial in insulin-naive patients with type 2 diabetes," *Lancet Diabetes Endocrinol.*, vol. 2, no. 11, pp. 885–893, Nov. 2014, doi: 10.1016/S2213-8587(14)70174-3.

[33] "Glucose Management Indicator (GMI): A New Term for Estimating A1C From Continuous Glucose Monitoring | Diabetes Care | American Diabetes Association." Accessed: Aug. 28, 2024. [Online]. Available: https://diabetesjournals.org/care/article/41/11/2275/36593/Glucose-Management-Indicator-GMI-A-New-Term-for

[34] A. El Fathi, M. Ganji, D. Boiroux, H. Bengtsson, and M. D. Breton, "Intermittent Control for Safe Long-Acting Insulin Intensification for Type 2 Diabetes: In-Silico Experiments," in *2023 IEEE Conference on Control Technology and Applications (CCTA)*, IEEE, 2023, pp. 534–539. Accessed: Aug. 23, 2024. [Online]. Available: https://ieeexplore.ieee.org/abstract/document/10253127/

[35] A. Schrijver, *Theory of Linear and Integer Programming*. John Wiley & Sons, 1998.